\documentstyle[twocolumn, prl, aps, epsf]{revtex}
\def\bea{\begin{eqnarray}}
\def\eea{\end{eqnarray}}
\begin{document}

\draft
\title{\bf $H_{c_3}$ for a thin-film superconductor 
with a ferromagnetic dot}

\author{Sa-Lin Cheng and H.A. Fertig}
\address{Department of Physics and Astronomy,
University of Kentucky,  Lexington, KY 40506-0055}
\date{\today}
\maketitle

\begin{abstract}
We investigate the effect of a ferromagnetic dot on a thin-film 
superconductor. We use a real-space method to solve the linearized
Ginzburg-Landau equation in order to find the upper critical field,
$H_{c_3}$. We show that $H_{c_3}$ is crucially dependent on dot 
composition and geometry,
and may be significantly greater than $H_{c_2}$.
$H_{c_3}$
is maximally enhanced when (1) the dot saturation magnetization is large, 
(2) the ratio of dot thickness to dot diameter is of order one, and 
(3) the dot thickness is large.
\end{abstract}

\section{INTRODUCTION}
Recent experiments\cite{martin,geoffroy,wernsdorfer,jaccard} 
involving thin-film superconductors
have investigated the effects of nanosize artificial pinning
centers in the form of ferromagnetic dots. Magnetic dots with
diameters on the order of 200 nm and thickness on the order of
40 nm were fabricated on a superconducting Nb film by electron
beam lithography. It was found that a regular array of magnetic
dots can dramatically influence the transport properties in the
presence of an applied magnetic field. In particular, the resistivity 
displayed minima at ``matching fields'' in which the number of flux 
quanta per dot was an integer.  Such effects are known to occur
in regular arrays of empty holes in thin-film superconductors
\cite{martin}.  These effects are believed to arise because empty 
holes can form effective pinning centers for multiple flux quanta 
vortices\cite{buzdin}, leading to particularly stable configurations 
at the matching fields. The strong pinning leading to such multiple 
vortices arises due to an enhancement of the order parameter near the 
edges of the hole, in a manner analogous to surface superconductivity
\cite{saint-james,degennes,tinkham}.

However, the physical situation for magnetic dots is significantly
different.  For empty holes, the superconducting order parameter
must have a vanishing derivative\cite{degennes,tinkham} at the vacuum/
superconductor interface.  In a magnetic field, the states satisfying 
the linearized Ginzburg-Landau equation and this boundary condition
turn out to have a maximum just inside the superconductor, leading to 
a magnetic field $H$ at which the superconducting order parameter may 
be non-zero near the surface while it vanishes in the bulk of the sample 
(i.e.,$H > H_{c_2}$).  This is very similar to surface superconductivity
\cite{tinkham}, and the maximum field at which superconducting order 
survives in the empty hole system is a direct analog of $H_{c_3}$ 
\cite{buzdin,tinkham}.

For magnetic dots, the strong field present inside the ferromagnet
supresses the superconducting order parameter, and in such situations
it is appropriate to adopt a boundary condition in which the order
parameter itself vanishes. This spoils the effect that leads to
surface superconductivity, and it is not at first obvious why magnetic
dots should support the relatively large supercurrents associated
with multiple vortices. However, the problem of a ferromagnetic
dot in a superconducting thin film has a dimension not present
in the empty hole analog: the magnetization and fringing magnetic
field of the dot itself. (Throughout this work, we will consider
only dots small enough to be treated as single domain ferromagnets.)
For dots with diameter $2R$ larger than their 
height $t$ (see Fig. 1) \cite{marmokous}, 
shape anisotropy dictates that the magnetization of the dot will lie
in the plane of the superconductor\cite{cullity} in the absence
of any external field. The application of a perpendicular magnetic field 
tilts the magnetization out of the plane of the dot, introducing a component 
to the dot fringing field that partially cancels the external applied flux 
passing through the superconducting film. This introduces a region just 
outside the dot in which the net field intensity is smaller than the applied 
field, allowing an enhancement of superconducting order.

\begin{figure}
\centerline{\epsfxsize=3.35in\epsfbox{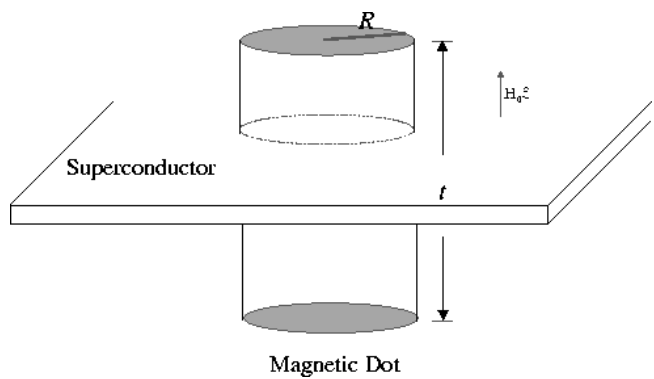}}
\vspace{3mm}
\caption{Magnetic dot with a radius $R$ and a thickness $t
(<2R)$ at the center of a thin-film superconductor. An external
field $H_0\hat z$ is applied through the sample, where $\hat z$ is the
direction parallel to the normal of the sample plane.}
\end{figure}

To demonstrate this effect, in this work we study the analog of 
$H_{c_3}$ in the presence of a single ferromagnetic dot with a diameter greater 
than its height ($2R>t$). To do this, we solve the linearized Ginzburg-Landau 
(GL) equation using a real-space method to be described below.
The resulting equations specify a maximum magnetic field $H_{c_3}(l)$
at which a non-zero superconducting order parameter may
be present for each value of the vorticity $l$.  
A typical example of our results is shown in Fig. 2. 
The form of this figure is easily understood if one
keeps in mind the analogy between the linearized GL
equation and the problem of electrons in a magnetic
field\cite{prange}.  In this analogy the vorticity $l$
plays the role of angular momentum, and it is well known
that, in the absence of a dot, the lowest-lying single particle 
states of a given $l$ in a magnetic field are localized near a radius 
$R_l = a_m\sqrt{2 l}$, where $a_m=\sqrt{\frac{\hbar c}{e^* H_0}}$ is the magnetic length,
$e^*$ the charge of the carriers, and $H_0$ is the applied
field.  While the presence of the dot and the
appropriate boundary condition changes the precise relation
between $l$ and $R_l$, it nevertheless remains generally
true that $R_l$ increases with $l$.
Thus, for small values of $l$, $H_{c_3}(l)$ is
supressed due to the boundary condition on the order parameter,
whereas for large $l$, $H_{c_3}(l) \rightarrow H_{c_2}$,
the value one expects in the absence of the dot.  For
intermediate values of $l$, one generically sees a peak
in $H_{c_3}(l)$, due to the fringing field effect described above.

\begin{figure}
\centerline{\epsfxsize=3.5in\epsfbox{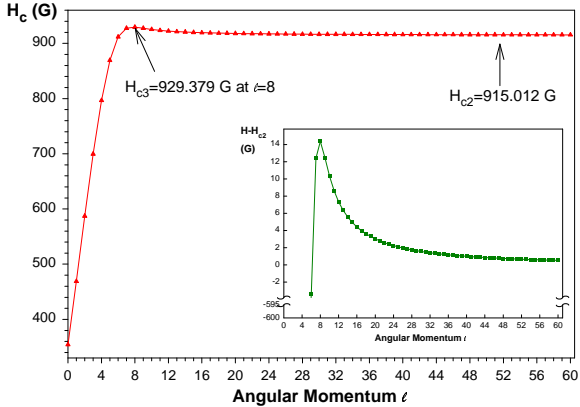}}
\caption{Magnetic field versus vorticity curve
for a Nb superconducting film with a Ni dot ($R$=100 nm and
$t$=40nm) at T=8.2 K. For large $l$, the field is unaffected by the
presence of the dot, giving $H_{c_2}$ = 915.012 G. The maximal $H_{c_3}$ (= 929.379 G)
occurs at $l$=8. The inset shows that the difference between $H_c$
and $H_{c_2}$ is maximal at $l$=8.}
\end{figure}

This peak value of $H_{c_3}(l)$ gives the maximum applied field in which
the thin film may sustain superconducting order, and is the analog
of $H_{c_3}$.  That it occurs at a finite value of $l$ indicates
that it is indeed true that magnetic dots support and presumably
pin multiple vortices.  It is interesting to note that for
appropriate parameters, this peak may become quite pronounced, and
that it may exceed the value of $H_{c_3} = 1.695 H_{c_2}$
that occurs for a simple infinite surface and represents the maximum
value possible in an empty hole\cite{buzdin}.  Such large 
values of $H_{c_3}$ occur when (1) the saturation magnetization
of the dot is large, (2) the height of the magnetic dot $t$ is
large, and (3) the dot diameter $2R$ is close to the height $t$.

This article is organized as follows.  In Section II we describe
our model of a ferromagnetic dot in a thin superconducting
film in detail and present the corresponding GL
equation for the system. Section III outlines our method for solving
the equation, and in Section IV we present our results.  We 
conclude in Section V with a summary.

\section{Model of a Ferromagnetic Dot in a Thin-Film Superconductor}

Consider a thin-film superconductor with a small
magnetic dot at its center (Fig. 1). The radius of the dot is
$R$, and the thickness is $t$. We assume that the dot is small
enough so that its magnetization density is uniform throughout
the dot; i.e., there are no domains. We wish to compute the
largest magnetic field $H_{c_3}$ for which the order parameter is
non-vanishing inside the superconductor.
        
The magnetization of the ferromagnet is maximal, but
its direction may vary. The orientational energy in any given
direction can be conveniently estimated if we approximate its
cylindrical shape as an ellipsoid whose semimajor axis length
is $c (=R)$ and semiminor axis length is $a (=t/2)$, as shown in
Fig. 3. 

\begin{figure}
\centerline{\epsfxsize=2.3in\epsfbox{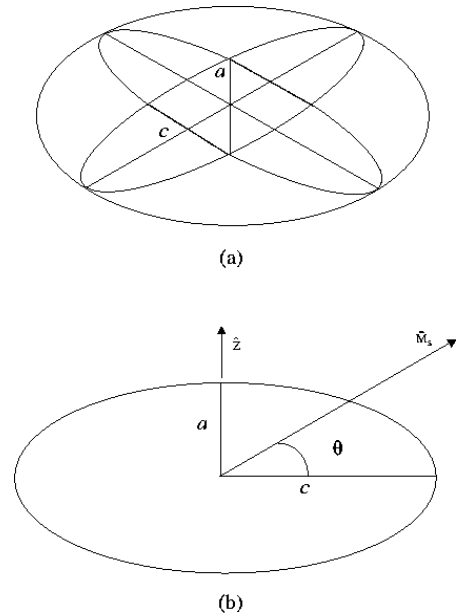}}
\vspace{3mm}
\caption{(a) Oblate spheroid with semimajor axis $c$ and
semiminor axis $a$. (b) $\hat z//a$. The angle between $M_s$ and the sample
plane is $\theta$.}
\end{figure}

 The energy for a given orientation of $\vec M$ 
(the dipole moment of the ferromagnet) in an ellipsoid is\cite{cullity}
\bea
F_M^0=\frac{1}{2}\Omega[(M_s\cos{\theta})^2N_c+(M_s\sin{\theta})^2N_a],
\label{Eq:eq1}
\eea
where $\Omega$ is the volume of ellipsoid, $M_s$ is the saturation
magnetization of the ferromagnet, and $N_a$ and $N_c$ are the
demagnetizing factors along $a$ and $c$. $\theta$ is the angle between $\vec M$
and the sample plane. Since $N_a>N_c$, $\sin{\theta}=0$ in the absence of an
applied magnetic field and the magnetization is perpendicular to
the $a$-axis, i.e., anywhere in the sample plane.

When there is an external field $H_0\hat z~(\hat z//a)$, we must
add $-\Omega\vec M\cdot H_0 \hat z$ to the dot energy. Therefore, the total 
magnetization energy of the dot is
\bea
F_M=-\Omega M_s H_0\sin{\theta}+F_M^0(\theta).
\label{Eq:eq2}
\eea                                                           
For a thin-film superconductor, the free energy may be
written as
\bea
F_S=F_N+\int_{r>R}d^2r[\alpha|\psi|^2+\frac{1}{2}\beta|\psi|^4+ \nonumber\\
 \frac{1}{2m}|(\frac{\hbar}{i}\nabla-\frac{e^*\vec A}{c})\psi|^2].                                                           
\label{Eq:eq3}
\eea
Since the superconductor is thin, we ignore any 
fields produced by supercurrents in the film.
The vector potential $\vec A$ then is the sum of the vector
potential due to the magnetic dot ($\vec {A_m}$) and the vector
potential due to the external field ($\vec {A_0}$): $\vec A=\vec {A_m}+\vec {A_0}$. 
In a uniform external field $H_0\hat z$, $\vec {A_0}$, in cylindrical coordinates,
is given by
\bea                                                            
\vec {A_0}=\frac{1}{2}H_0r\hat \varphi.
\label{Eq:eq4}
\eea
Furthermore, for a thin-film superconductor, magnetic field components
parallel to the film have no effect on the order
parameter\cite{tinkham}. Thus, we may write
\bea
\vec {A_M}(r)=M_st\sin{\theta}a_0(\frac{r}{R})\hat \varphi,
\label{Eq:eq5}
\eea                                                            
where
\bea
a_0(x)\equiv\int_{-\pi}^\pi d\varphi\frac{\cos {\varphi}}
 {[x^2-2x\cos{\varphi}+1]^{1/2}},
\label{Eq:eq6}
\eea                                                           
which is an elliptical integral \cite{com}.  
Therefore, the total free energy of the system is $F=F_S+F_M$, or
\begin{eqnarray}
F=F_N+\int_{r>R}d^2r[\alpha|\psi|^2+\frac{1}{2}\beta|\psi|^4+ \nonumber\\
\frac{1}{2m}|(\frac{\hbar}{i}\nabla-\frac{e^*\vec A}{c})\psi|^2]   
 -\Omega M_s H_0\sin{\theta}+ \nonumber\\
\frac{1}{2}\Omega[(M_s\cos{\theta})^2N_c+(M_s\sin{\theta})^2N_a].                                                          
\label{Eq:eq7}
\end{eqnarray}        
We need to minimize the total free energy with respect
to $\psi$ and $\theta$ (or $\sin{\theta}$). After minimizing with respect to $\sin{\theta}$, we obtain
\begin{eqnarray}
-\frac{e^*}{2mc}\int_{r>R}d^2r[\psi^*(\frac{\hbar}{i}\nabla-\frac{e^*\vec A}{c})\psi+ \nonumber\\
\psi(-\frac{\hbar}{i}\nabla-\frac{e^*\vec A}{c})\psi^*]\cdot M_sta_0(\frac{r}{R})\hat\varphi\nonumber\\
=-\Omega M_s H_0+M_s^2\Omega(N_a-N_c)\sin{\theta}.
\label{Eq:eq8}                               
\end{eqnarray}
To find $H_{c_3}$, we consider $|\psi|<<1$ and so drop terms involving
$\psi$. Then we have
\bea
\sin{\theta}=\frac{H_0}{M_s(N_a-N_c)}.
\label{Eq:eq9}                                                           
\eea
After minimizing the free energy with respect to $\psi$, we obtain
\bea
\frac{1}{2m}(\frac{\hbar}{i}\nabla-\frac{e^*\vec A}{c})^2\psi=|\alpha|\psi, \alpha<0,                                                           
\label{Eq:eq10}
\eea
where $\alpha\propto T-T_c$. Eq. (10) is the linearized Ginzburg-Landau
equation.  It should be kept in mind that although the
equation is linear in $\psi$, it is {\it non-linear} in
$\sin{\theta}$.  This means our solution of Eq. (10) requires
a level of self-consistency usually absent in solving the linearized
GL equation, which we describe below.

Eq. (7) essentially defines our model of the film-dot system,
and Eqs. (9) and (10) are what need to be solved to obtain the
critical field of the system.  In the next section we present
some details describing how this is done.

\section{Numerical Solution of Linearized Ginzburg-Landau Equation}

We need to find the largest value of $H_0$ (i.e., $H_{c_3}$)
for which the eigenvalue equation (Eq. 10) has a non-trivial solution
with the boundary condition
\bea
\psi(r=R)=0.
\label{Eq:eq11}
\eea                                                           
Due to circular symmetry, we can write the solution to Eq. (10)
in the form
\bea
\psi(\vec r)=f(r)e^{il\varphi}, 
\label{Eq:eq12}                   
\eea
where $l$ is an integer. The orbital number, $l$, corresponds to the 
vorticity of
the solution for the superconducting order parameter.
Next, we scale out length by defining
\bea
\rho\equiv\frac{r}{a_m},
\label{Eq:eq13}    
\eea
where the magnetic length $a_m$ is
\bea
a_m=(\frac{\hbar c}{e^*H_0})^{1/2} .
\label{Eq:eq14}
\eea
Applying Eq. (12) and Eq. (13), Eq. (10) becomes           
\bea
[-\frac{1}{\rho}\frac{d}{d\rho}\rho\frac{d}{d\rho}+(\frac{l}{\rho}-\tilde A(\rho))^2]f(\rho)=\varepsilon f(\rho),                                                         
\label{Eq:eq15}
\eea
where
\bea
\tilde A(\rho)=\frac{1}{2}\rho+M_st\frac{\sin{\theta}}{a_mH_0}a_0(\frac{a_m\rho}{R})
\label{Eq:eq16}
\eea                                                      
and                
\bea                                       
\varepsilon=\frac{2m|\alpha|}{\hbar^2}(\frac{\hbar c}{e^*H_0}).
\label{Eq:eq17}
\eea
The smallest eigenvalue of Eq. (\ref{Eq:eq15}) corresponds to the
largest $H_0$ for which there is a non-vanishing order parameter.
Once we find $\varepsilon_0$, we can directly write
\bea
H_{c_3}=\frac{2mc|\alpha|}{\hbar e^*\varepsilon_0}. 
\label{Eq:eq18}                                                          
\eea
To solve this eigenvalue problem, we used a real space
method as follows:
\begin{itemize}
  \item[(1)] Guess $H_0=H_{c_3}$. Notice that $H_0$ enters explicitly in the
vector potential and $\sin{\theta}$, and thus cannot be scaled out as
would be the case for an empty hole. 
  \item[(2)] Define a set of N points {$\rho_1,\rho_2, \rho_3,\ldots,\rho_N$} and $\rho_0\equiv\frac{R}{a}$.
  \item[(3)] Turn the derivatives into differences, thereby transforming
the differential equation into a {\it difference equation}.
  \item[(4)] Set up the column vector 
$$
\left[
\begin{array}{c}
f_1\\f_2\\f_3\\ \vdots\\f_N
\end{array}
\right],
$$
and define $\Delta X_{n+1/2}=\rho_{n+1}-\rho_n, \Delta X_{n-1/2}=\rho_n-\rho_{n-1}$,
turning the differential equation into a matrix equation:
\bea
\left[  
 \begin{array}{ccccc}
 \beta_1 & \alpha_1 & 0 & 0 &  \cdots \\
 \gamma_2 & \beta_2 & \alpha_2 & 0 &  \cdots\\
 0 & \gamma_3 & \beta_3 & \alpha_3 &  \cdots\\
 0 & \ddots & \ddots & \ddots & \ddots\\ 
 0 & \ddots & \ddots & \ddots & \ddots
 \end{array}
 \right] 
\left[
\begin{array}{c}
f_1\\f_2\\f_3\\ \vdots\\f_N
\end{array}
\right]=\varepsilon_0
\left[
\begin{array}{c}
f_1\\f_2\\f_3\\ \vdots\\f_N
\end{array}
\right]
\label{Eq:eq19}
\eea  
where
\bea
\alpha_n = -\frac{8}{(\Delta X_{n+1/2}+\Delta X_{n-1/2})^3}\Delta X_{n-1/2}- \nonumber\\
            \frac{1}{\rho_n(\Delta X_{n+1/2}+\Delta X_{n-1/2})},
\eea
\bea
\beta_n = \frac{8}{(\Delta X_{n+1/2}+\Delta X_{n-1/2})^2}+ \nonumber\\
         (\frac{l}{\rho_n}-\tilde A(\rho_n))^2,
\eea
and
\bea
\gamma_n = -\frac{8}{(\Delta X_{n+1/2}+\Delta X_{n-1/2})^3}\Delta X_{n+1/2}+ \nonumber\\
           \frac{1}{\rho_n(\Delta X_{n+1/2}+\Delta X_{n-1/2})}.
\eea                      
              
  \item[(5)] Diagonalize the matrix to find the lowest eigenvalue, $\varepsilon_0$,
which gives the highest magnetic field 
$H^*=\frac{2mc|\alpha|}{\hbar e^*\varepsilon_0}$.    
  \item[(6)] Finally, one must check if the solution is
self-consistent. 
If $H^*[\varepsilon_0]$ equals the guessed $H_0$ in step (1), 
then $H_0 = H_{c_3}$.
However,  if $H^*[\varepsilon_0]$  is not equal to the initial guess, then set
$H^*\to H_0$ and return to (1).  The element of self-consistency
arises because one must determine the orientation of the
dot magnetization, $\theta$, in the field $H_{c_3}$.
Note that if the field $H_{c_3}$ is large enough, then
$\sin{\theta}=1$ and the magnetization is fully parallel
to the applied field; in this situation Eq. (9) is not
appropriate (except precisely at $H_0/(M_s(N_a-N_c)) = 1$),
and we do not need to iterate the equations.
\end{itemize}

\section{Results}

One of the interesting results that was found in this study
is that $H_{c_3}$ may be very large for this system.
In order to examine when this happens, we studied how the
variables, $M_s$, $t$ and $a/c$ (or $N_a-N_c$) affect $H_{c_3}$. First, we
fixed $t$ and $a/c$ to compute $H_{c_3}$ for different values of $M_s$.
Clearly, we expect a large $M_s$ to create large fringing
field, capable of cancelling large applied fields near
the dot.  Presumably this will lead to large values of
$H_{c_3}$.  Fig. 4 illustrates this for
a 160 nm thick dot with $a/c (=t/(2R))$ = 0.9 (or $N_a=
4\pi~\times~$0.361472 and $N_c=4\pi~\times~$0.305689) in a Nb film at T=8.2 K,
with Ni ferromagnetic dots ($M_s$=509 G).
(These parameters correspond to those of the magnetic dots
studied in Ref. \onlinecite{martin}.)  For comparison, 
we also show $H_{c_3}$ for a Dy dot ($M_s$=2920 G) with otherwise
the same parameters of the system.  The enhancement of
$H_{c_3}$ for the larger value of $M_s$ is quite apparent.

\begin{figure}
\centerline{\epsfxsize=3.5in\epsfbox{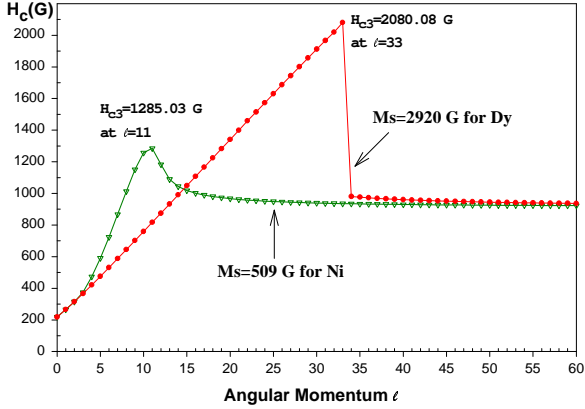}}
\caption{Magnetic field versus vorticity curves
for a Nb superconducting film with a dot at T=8.2 K. The dot
size is fixed: $t$ =160 nm and $t/(2R)$ = 0.9. If the dot material
is Dy ($M_s$=2920 G), $H_{c_3}$ = 2080.08 G at $l$=33. However, if the
material is Ni ($M_s$=509 G), $H_{c_3}$ = 1285.03 G at $l$=11.}
\end{figure}

We also note in this figure and several that follow that
there is an apparent jump in $H_{c_3}(l)$ vs. $l$.  If
$l$ were a continuous variable this would not be a discontinuous
jump but rather a continuous (albeit sharp) drop in $H_{c_3}(l)$.
Nevertheless, the sharp behavior is a direct result of the
nonlinearity of the equations in $\sin{\theta}$.  The behavior
represents a sharp crossover as a function of $l$ in which the
superconducting order parameters are localized relatively close to the 
dot and ones in which they are further away; in the latter case 
the dot potential
is a relatively weak perturbation on the result in the absence
of a dot.

Next, we fixed $M_s$ and $a/c$ to study the influence of
dot thickness $t$ on $H_{c_3}$.  For large values of $t$ one
again expects the field generated by the dot can cancel a relatively large
external field, leading to an enhancement of $H_{c_3}$.
One can imagine a situation in which both the film and 
the dot thickness $t$ are the same and are varied together.
The results reported here will apply provided the superconductor
is effectively two-dimensional -- i.e., the coherence length
must be larger than $t$.  For many interesting materials
that may not be possible; however, it is quite possible and
often appropriate to consider systems in which the dot
thickness is {\it different} than that of the film.  This
is the situation depicted in Fig. 1.
Fig. 5 illustrates $H_{c_3}$ for a Ni dot with
$a/c$=0.9 in a Nb film at T=8.2K. As expected,
thicker dots (bigger $t$) indeed yield higher values of $H_{c_3}$.

\begin{figure}
\centerline{\epsfxsize=3.5in\epsfbox{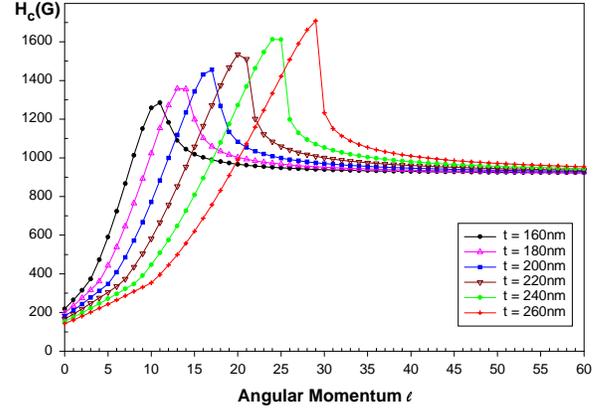}}
\caption{Magnetic field versus vorticity curves
for a Nb superconducting film with a Ni dot at T=8.2 K. The dot
thickness varies from 160 nm to 260 nm, while keeping the value
of $t/(2R)$ fixed. $H_{c_3}$ increases with $t$.}
\end{figure}

Finally, we fixed $M_s$ and $t$ and 
considered the effect of $a/c$ on $H_{c_3}$. The aspect
ratio of the dot is relevant because it enters into the
demagnetizing factors $N_a$ and $N_c$. Physically, if
$a/c$ is close to one, the dot should approach a limit in
which it is relatively easy to tip the magnetization
out of the plane\cite{com2}.  This again maximizes the fringing field
available to cancel the applied field.  Fig. 6 illustrates
$H_{c_3}$ for $a/c$=0.8 and $a/c$=0.9 
(or $N_a-N_c$=1.4860487 and $N_a-N_c$=0.7009889, respectively),
and demonstrates that greater values of $a/c$ (or smaller values of
$N_a-N_c$) give higher values of $H_{c_3}$.

\begin{figure}
\centerline{\epsfxsize=3.5in\epsfbox{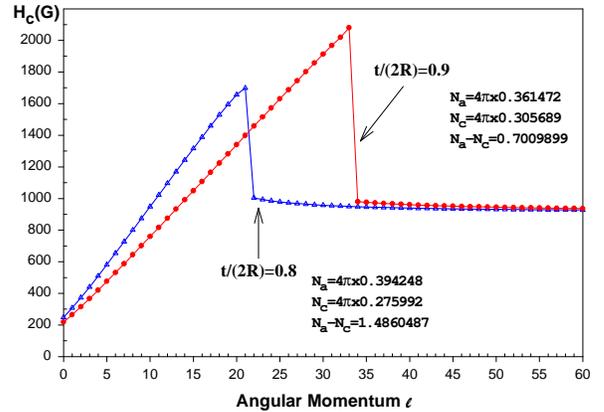}}
\caption{Magnetic field versus vorticity curves
for a Nb superconducting film with a Dy dot at T=8.2 K. The
dot's thickness is fixed at $t$ = 160 nm, but the radius varies:
$t/(2R)$ = 0.9 (or $N_a-N_c$=0.7009899) and $t/(2R)$ = 0.8 (or
$N_a-N_c$=1.4860487). The curve for $t/(2R)$ = 0.9 shows a greater
$H_{c_3}$ than that for $t/(2R)$ = 0.8.}
\end{figure}

\section{Summary}

        In this work, we studied $H_{c_3}$ for magnetic dots in a
thin-film superconductor. To find $H_{c_3}$, we used a real-space
method to solve the linearized Ginzburg-Landau equation. We
showed that the enhancement of the order parameter crucially
involves the shape anisotropy of the magnetic dot and is qualitatively
different from that for empty holes. We found the enhancement
to be maximal when (1) the dot saturation magnetization $M_s$
is large,  (2) the dot thickness is large, and (3) the value
of $t/(2R)$ of the dot is close to 1, for which the demagnetizing
parameters $N_a$ and $N_c$ are of comparable magnitude.

\vspace{5mm}
\centerline{\bf ACKNOWLEDGMENTS}
\vspace{5mm}
The authors thank J.J. Palacios for helpful discussions in the
early stages of this work.
This work was supported by NSF Grant Nos. DMR98-70681 and PHY94-07194,
and the Research Corporation.  HAF thanks the ITP at UC Santa Barbara
for its hospitality.


\begin{references}
\bibitem{com} This form
for the dot vector potential is easily derived when
one notes that the magnetic field produced by (the $z$-component of)
a uniformly magnetized cylinder is the same as that of a current
sheet around the side of the cylinder.
\bibitem{com2} Note that if $a/c$ is greater than one, then the
equilibrium direction of the magnetization will be along the
cylinder axis rather than in the plane.  In this case fringing
fields will be present and disrupt the superconducting order
even in the absence of an applied field.  This situation was
considered in Ref.\onlinecite{marmokous} above, and we do not
address it in our present work.
\end{references}
\end{document}